# Role of entanglement in two-photon imaging

Ayman F. Abouraddy, Bahaa E. A. Saleh[†], Alexander V. Sergienko, and

Malvin C. Teich

*Quantum Imaging Laboratory,[‡] Departments of Electrical & Computer Engineering and*

*Physics, Boston University, 8 Saint Mary's Street, Boston, MA 02215-2421*

The use of entangled photons in an imaging system can exhibit effects that cannot be mimicked by any other two-photon source, whatever the strength of the correlations between the two photons. We consider a two-photon imaging system in which one photon is used to probe a remote (transmissive or scattering) object, while the other serves as a reference. We discuss the role of entanglement versus correlation in such a setting, and demonstrate that entanglement is a prerequisite for achieving distributed quantum imaging.

[†] Electronic address: besaleh@bu.edu

[‡] URL: http://www.bu.edu/qil

A quantum two-particle system in an entangled state exhibits effects that cannot be attained with any classically correlated system, no matter how strong the correlation. The difference between entanglement and classical correlation lies at the heart of Bell-type experiments. In this paper, we consider general configurations for two-photon imaging, with the goal of discerning the true role of entanglement in contradistinction to correlation. We identify the unique advantage a source of entangled photon pairs has over a source of classically correlated photon pairs in this context. Although imaging with entangled photons generated by spontaneous parametric down-conversion has been proposed [1] and examined in some simple settings [2,3], it has not been clear whether such experiments may, in principle, be reproduced using a classically correlated source.

*Single-photon imaging.*— To set the stage for our analysis of two-photon imaging systems, we begin with a brief summary of the basic equations governing single-photon imaging. It is well known [4] that a single photon traveling through an optical system exhibits all the phenomena of diffraction, interference, and imaging that are familiar in classical optics. One needs only to repeat the single-photon experiment and accumulate observations over a sufficiently large ensemble. Consider, for example, a thin planar single-photon source described by the pure state

$$|\Psi\rangle = \int d\mathbf{x}\, \varphi(\mathbf{x}) |1_\mathbf{x}\rangle, \tag{1}$$

where $|1_\mathbf{x}\rangle = \frac{1}{(2\pi)^2} \int d\mathbf{k}\, e^{i\mathbf{k}\cdot\mathbf{x}} |1_\mathbf{k}\rangle$ is a position representation of the single-photon state in terms of the familiar Fock state $|1_\mathbf{k}\rangle$ of the mode $\mathbf{k}$, and the state probability amplitude $\varphi(\mathbf{x})$ is normalized such that $\int d\mathbf{x}\, |\varphi(\mathbf{x})|^2 = 1$. Assume now that this state is transmitted through a linear optical system, described by an impulse response function $h(\mathbf{x}_1, \mathbf{x})$, where $\mathbf{x}$ and $\mathbf{x}_1$ are the transverse coordinates on the input and output planes, respectively, as illustrated in Fig. 1. If a photon arrives in the output plane, the probability density of its registration at a position $\mathbf{x}_1$ is [5]

$$p(\mathbf{x}_1) \propto \left| \int d\mathbf{x}\, \varphi(\mathbf{x}) h(\mathbf{x}_1, \mathbf{x}) \right|^2, \qquad (2)$$

where $\int d\mathbf{x}_1 p(\mathbf{x}_1) = 1$. Equation (2) is the familiar relation describing a coherent optical system [6].

If the single-photon source is, instead, in a mixed state described by a density operator

$$\hat{\rho} = \iint d\mathbf{x}\, d\mathbf{x}'\, \gamma(\mathbf{x}, \mathbf{x}') |1_\mathbf{x}\rangle \langle 1_{\mathbf{x}'}|, \qquad (3)$$

where $\int d\mathbf{x}\, \gamma(\mathbf{x}, \mathbf{x}) = 1$ and $\gamma(\mathbf{x}, \mathbf{x}') = \gamma^*(\mathbf{x}', \mathbf{x})$, then the photon probability density is

$$p(\mathbf{x}_1) \propto \iint d\mathbf{x}\, d\mathbf{x}'\, \gamma(\mathbf{x}, \mathbf{x}') h(\mathbf{x}_1, \mathbf{x}) h^*(\mathbf{x}_1, \mathbf{x}'), \qquad (4)$$

which is the familiar equation describing a partially coherent optical system [6]. It therefore follows that the behavior of an optical system with a single-photon source in an arbitrary mixed state is analogous to that of the same system illuminated with partially coherent light. The incoherent limit is attained when $\gamma(\mathbf{x}, \mathbf{x}') = \gamma(\mathbf{x}) \delta(\mathbf{x} - \mathbf{x}')$ whereas the coherent limit emerges when $\gamma$ is factorizable in the form $\gamma(\mathbf{x}, \mathbf{x}') = \varphi(\mathbf{x}) \varphi^*(\mathbf{x}')$, whereupon Eq. (2) is recovered.

*Two-photon imaging.*— We now consider a planar source that emits light in the pure two-photon state

$$|\Psi\rangle = \iint d\mathbf{x}\, d\mathbf{x}'\, \varphi(\mathbf{x}, \mathbf{x}') |1_\mathbf{x}, 1_{\mathbf{x}'}\rangle, \qquad (5)$$

where $\iint d\mathbf{x}d\mathbf{x}'|\varphi(\mathbf{x},\mathbf{x}')|^2 = 1$ and $|1_\mathbf{x},1_{\mathbf{x}'}\rangle$ is the two-photon state at transverse coordinates $\mathbf{x}$ and $\mathbf{x}'$. For full generality, we assume that the emitted photons are transmitted through different linear optical systems with impulse response functions $h_1(\mathbf{x}_1,\mathbf{x})$ and $h_2(\mathbf{x}_2,\mathbf{x}')$, where $\mathbf{x}_1$ and $\mathbf{x}_2$ are transverse coordinates on the output planes of the two systems, as depicted in Fig. 2. There are now many options for the placement of objects as well as many observation schemes. A single object may be placed in either system, or objects may be placed in both systems. The photon coincidence at $\mathbf{x}_1$ and $\mathbf{x}_2$ may be recorded, yielding the joint probability density $p(\mathbf{x}_1,\mathbf{x}_2)$ [5]

$$p(\mathbf{x}_1,\mathbf{x}_2) \propto \left|\iint d\mathbf{x}d\mathbf{x}'\varphi(\mathbf{x},\mathbf{x}')h_1(\mathbf{x}_1,\mathbf{x})h_2(\mathbf{x}_2,\mathbf{x}')\right|^2. \tag{6}$$

This function is the fourth-order correlation function (also called the coincidence rate) $G^{(2)}(\mathbf{x}_1,\mathbf{x}_2)$ [5, 7].

*Two distinct single-photon probability densities.*— We now consider two *distinct* single-photon probability density functions associated with this two-photon system. The first is the probability density of observing a photon at $\mathbf{x}_1$, regardless of whether the other photon is detected (or even whether there is another photon). This is called the single-photon probability density $p_1(\mathbf{x}_1)$. The second is the probability density of observing a photon at $\mathbf{x}_1$ at the output of system 1 *and* a photon at *any* location ($-\infty < \mathbf{x}_2 < \infty$) at the output of system 2. We call this the marginal probability density $\bar{p}_1(\mathbf{x}_1)$ [8]. We similarly define $p_2(\mathbf{x}_2)$ and $\bar{p}_2(\mathbf{x}_2)$. We proceed to demonstrate that, remarkably, these two probability densities are *not* always identical.

The *single-photon probability density* $p_j(\mathbf{x}_j)$, $j = 1, 2$, is obtained by taking the trace over the subspace of the other photon. Each photon, considered separately from the other, is in fact in a mixed state described by the density operator provided in Eq. (3):

$$\hat{\rho}_j = \iint d\mathbf{x} d\mathbf{x}' \gamma_j(\mathbf{x}, \mathbf{x}') |1_\mathbf{x}\rangle\langle 1_{\mathbf{x}'}|, \quad j = 1, 2, \tag{7}$$

where $\gamma_1(\mathbf{x}, \mathbf{x}') = \int d\mathbf{x}'' \varphi(\mathbf{x}, \mathbf{x}'') \varphi^*(\mathbf{x}', \mathbf{x}'')$ and $\gamma_2(\mathbf{x}, \mathbf{x}') = \int d\mathbf{x}'' \varphi(\mathbf{x}'', \mathbf{x}) \varphi^*(\mathbf{x}'', \mathbf{x}')$. The probability density of a photon registration at the output of system $j$ is then given by

$$p_j(\mathbf{x}_j) \propto \iint d\mathbf{x} d\mathbf{x}' \gamma_j(\mathbf{x}, \mathbf{x}') h_j(\mathbf{x}_j, \mathbf{x}) h_j^*(\mathbf{x}_j, \mathbf{x}'), \quad j = 1, 2. \tag{8}$$

This expression is similar to Eq. (4) for a single photon in a mixed state. It follows that the optics of a single photon from a two-photon source that is in a pure state exhibits the behavior of a partially coherent system.

In contrast to Eq. (8), the *marginal probability density* of detecting a photon at detector D$_1$ is given by

$$\bar{p}_1(\mathbf{x}_1) = \int d\mathbf{x}_2 \, p(\mathbf{x}_1, \mathbf{x}_2). \tag{9}$$

This is the probability of observing one photon at $\mathbf{x}_1$, at the output of system 1, and another at the output of system 2 at *any* location. The detector at system 2 may then be termed a 'bucket' detector since it does not register the arrival location of the photon. In this conception, the bucket detector can serve as a 'gating' signal for a scanning detector at the output of system 1. The marginal probability density $\bar{p}_2(\mathbf{x}_2)$ may be similarly defined and measured.

Based on classical probability theory one would intuitively expect that $p_j(\mathbf{x}_j)$ would be equal to $\bar{p}_j(\mathbf{x}_j)$. This is not always the case, however. Indeed, in some situations, measurement of $\bar{p}_j(\mathbf{x}_j)$ at the output of one system can be used to extract information about an object placed in the other system.

*Probability densities for a non-entangled source.*— Let us examine a few special cases. We consider a non-entangled two-photon source in which the state probability amplitude is factorizable, i.e., $\varphi(\mathbf{x},\mathbf{x}') = \varphi_1(\mathbf{x})\varphi_2(\mathbf{x}')$. The joint probability density in this case is also factorizable. There is nothing to be gained by measuring the joint probability density (coincidence rate) since all information is contained in the single-photon probability densities (singles rates). The photon arrivals are independent, and each is governed by a coherent imaging system [5,9]. Moreover,

$$\bar{p}_j(\mathbf{x}_j) = p_j(\mathbf{x}_j), j = 1, 2, \tag{10}$$

so that this factorizable state, therefore, does not permit the transfer of information in one system to the other.

*Probability densities for an entangled source.*— Consider now an entangled [10] two-photon source described by the state probability amplitude $\varphi(\mathbf{x},\mathbf{x}') = \varphi(\mathbf{x})\delta(\mathbf{x}-\mathbf{x}')$, in which case the two-photon state is

$$|\Psi\rangle = \int d\mathbf{x}\,\varphi(\mathbf{x})|1_\mathbf{x},1_\mathbf{x}\rangle, \tag{11}$$

where the reduced density operators of the individual photons are $\hat{\rho}_1 = \hat{\rho}_2 = \int d\mathbf{x}|\varphi(\mathbf{x})|^2|1_\mathbf{x}\rangle\langle 1_\mathbf{x}|$, so that

$$p(\mathbf{x}_1,\mathbf{x}_2) \propto \left|\int d\mathbf{x}\,\varphi(\mathbf{x})h_1(\mathbf{x}_1,\mathbf{x})h_2(\mathbf{x}_2,\mathbf{x})\right|^2, \tag{12a}$$

$$p_j(\mathbf{x}_j) \propto \int d\mathbf{x}|\varphi(\mathbf{x})|^2|h_j(\mathbf{x}_j,\mathbf{x})|^2, \quad j=1,2, \tag{12b}$$

$$\bar{p}_j(\mathbf{x}_j) \propto \iint d\mathbf{x}\,d\mathbf{x}'\,\varphi(\mathbf{x})\varphi^*(\mathbf{x}')g_k(\mathbf{x},\mathbf{x}')h_j(\mathbf{x}_j,\mathbf{x})h_j^*(\mathbf{x}_j,\mathbf{x}'), \quad j,k=1,2;\ j\neq k, \tag{12c}$$

with $g_k(\mathbf{x}, \mathbf{x}') = \int d\mathbf{x}'' h_k(\mathbf{x}'', \mathbf{x}) h_k^*(\mathbf{x}'', \mathbf{x}')$, $k = 1, 2$. Evidently $\bar{p}_j(\mathbf{x}_j) \neq p_j(\mathbf{x}_j)$, $j = 1, 2$.

Whereas the expression in Eq. (12b) is similar to that for an incoherent optical system, Eq. (12c) is similar to that for a partially coherent optical system. Moreover, measurement of $\bar{p}_1(\mathbf{x}_1)$ contains information about the system function $h_2(\mathbf{x}_2, \mathbf{x})$ via the function $g_2(\mathbf{x}, \mathbf{x}')$.

The most accessible entangled two-photon source is based on the process of spontaneous parametric down-conversion (SPDC) from a nonlinear crystal [11]. For a monochromatic pump, assuming the down-converted beams are filtered with narrowband spectral filters, the state probability amplitude for SPDC is given by [5]

$$\varphi(\mathbf{x}, \mathbf{x}') \propto \int d\mathbf{x}'' E_p(\mathbf{x}'') \xi(\mathbf{x} - \mathbf{x}'', \mathbf{x}' - \mathbf{x}''), \tag{13}$$

where $E_p(\mathbf{x})$ is the pump field at the input of the crystal and $\xi(\mathbf{x}, \mathbf{x}')$ is a phase-matching function that depends on the crystal parameters. In the limit $\xi(\mathbf{x}, \mathbf{x}') \to \delta(\mathbf{x})\delta(\mathbf{x}')$, the state function reduces to $\varphi(\mathbf{x}, \mathbf{x}') = E_p(\mathbf{x})\delta(\mathbf{x} - \mathbf{x}')$, which corresponds to an entangled state. This can be achieved by reducing the thickness of the crystal. In this limit, the joint probability density, the single-photon probability density, and marginal probability density, which are given by Eqs. (6), (8), and (9) respectively, yield Eqs. (12a) - (12c). The spatial coherence properties of this source have been studied extensively [12]. Increasing the crystal thickness eventually leads to the limit of the factorizable state.

*Distributed quantum imaging.*— To demonstrate the utility of measuring the marginal probability density $\bar{p}_2(\mathbf{x}_2)$, consider the following scenario. Let system 1 comprise a scattering object (see Fig. 2). The scattered radiation impinges on detector $D_1$. Such a system cannot by itself yield an image of the spatial distribution of the scattering object. However, if $D_1$ is a bucket detector that gates the photon arrival registered by scanning $D_2$, one is able to form a high quality image using this two-photon scheme, as long as the scattering object is illuminated by one photon at a time. Using SPDC, simple experiments along these lines have been carried out [2,3].

However, the origin of this distributed quantum-imaging phenomenon has not been adequately set forth heretofore. Indeed, it was stated in Ref. [3] that "it is possible to imagine some type of classical source that could partially emulate this behavior". To obtain a better understanding of this effect we consider a two-photon source in an arbitrary mixed state. Can similar results be obtained by employing a two-photon source that exhibits classical statistical correlations but not entanglement? That is, can distributed quantum-imaging be achieved without entanglement?

To answer this question we take a mixed state that exhibits the strongest possible classical correlations, i.e., one for which

$$\hat{\rho} = \int d\mathbf{x}\, \gamma(\mathbf{x}) |1_\mathbf{x}, 1_\mathbf{x}\rangle\langle 1_\mathbf{x}, 1_\mathbf{x} |, \tag{14}$$

with $\int d\mathbf{x}\, \gamma(\mathbf{x}) = 1$. This state represents a superposition of photon-pair emission probabilities from various locations $\mathbf{x}$. (In contrast, the entangled state in Eq. (11) represents a superposition of probability amplitudes.) The density operators of each photon taken individually are $\hat{\rho}_1 = \hat{\rho}_2 = \int d\mathbf{x}\, \gamma(\mathbf{x}) |1_\mathbf{x}\rangle\langle 1_\mathbf{x}|$, which are identical to those of the entangled source if $\gamma(\mathbf{x}) = |\varphi(\mathbf{x})|^2$. In this case, then,

$$p(\mathbf{x}_1, \mathbf{x}_2) \propto \int d\mathbf{x}\, \gamma(\mathbf{x}) |h_1(\mathbf{x}_1, \mathbf{x})|^2 |h_2(\mathbf{x}_2, \mathbf{x})|^2, \tag{15a}$$

$$p_j(\mathbf{x}_j) \propto \int d\mathbf{x}\, \gamma(\mathbf{x}) |h_j(\mathbf{x}_j, \mathbf{x})|^2, \quad j = 1, 2, \tag{15b}$$

$$\bar{p}_j(\mathbf{x}_j) \propto \int d\mathbf{x} \bar{\gamma}_k(\mathbf{x}) |h_j(\mathbf{x}_j, \mathbf{x})|^2, \quad j,k = 1,2, \quad j \neq k, \tag{15c}$$

where $\bar{\gamma}_k(\mathbf{x}) = \gamma(\mathbf{x}) \int d\mathbf{x}' |h_k(\mathbf{x}', \mathbf{x})|^2$, $k = 1, 2$. The result in Eq. (15b) is similar to that in Eq. (12b), i.e., that of an incoherent optical system, since the reduced density operators are identical. The distinction is that Eq. (15c) has the form of an incoherent optical system whereas Eq. (12c) has the form of a partially coherent optical system. For the special case of a shift-invariant (isoplanatic) system, in which case $h_j(\mathbf{x}_j, \mathbf{x})$ is a function of $\mathbf{x}_j - \mathbf{x}$, then $\bar{\gamma}_k(\mathbf{x}) \propto \gamma(\mathbf{x})$, $k = 1, 2$, and Eq. (15c) becomes identical to Eq. (15b). As a result of Eqs. (15a-c), the distributed quantum-imaging scheme truly requires entanglement in the source and cannot be achieved using a classical source with correlations but without entanglement. There is, of course, a continuous transition between these two extremes, so that partial distributed quantum imaging is possible as entanglement enters the mixed state.

To appreciate the consequences of the imaging formulas given in Eqs. (4), (12a-c), and (15a-c), consider $N$ weak scatterers embedded in one of the optical systems ($h$ in the single-photon case or $h_1$ in the two-photon case). The impulse response function of the system including the scatterers is then given by:

$$h(\mathbf{x}_1, \mathbf{x}) = h_o(\mathbf{x}_1, \mathbf{x}) + \sum_{j=1}^{N} \varepsilon_j h^{(j)}(\mathbf{x}_1, \mathbf{x}_j), \tag{16}$$

where $h_o(\mathbf{x}_1, \mathbf{x})$ represents the system in the absence of the scatterers, $h^{(j)}(\mathbf{x}_1, \mathbf{x}_j)$ is the impulse response function of the system following a scatterer at location $\mathbf{x}_j$, and $\varepsilon_j$ is the strength of scatterer $j$. We assume that the system in which the scatterers are contained is inaccessible. By substituting Eq. (16) in Eqs.(4), (12a-c), and (15a-c), it turns out that in the one-photon case the image of the scatterers is in general blurred by a distribution that depends on $h^{(j)}(\mathbf{x}_1, \mathbf{x}_j)$, as expected. In the entangled-two-photon case, however, one can always select an $h_2(\mathbf{x}_2, \mathbf{x}')$ (see Fig. 2) such that the combination with $h^{(j)}(\mathbf{x}_1, \mathbf{x}_j)$ yields a diffraction limited imaging system for each scattering plane. The correlated-two-photon case, on the other hand, offers no such benefit.

*Conclusion.*— We have considered a distributed quantum imaging system in which one photon is used to probe a remote transmissive or scattering object, while the other serves as a reference. A high-spatial-resolution detector scans the arrival position of the reference photon, while a bucket detector (which need have no spatial resolution) registers the photon scattered by the object. The scanned reference detector is gated by the photoevent registered by the bucket detector, and an image is formed. We have shown that if the two-photon source is in an entangled state, the imaging is, in general, partially coherent, and can possibly be fully coherent. On the other hand, if the source emits unentangled, but classically-correlated photon pairs, then the imaging is incoherent.

We thank Zachary Walton for valuable discussions. This work was supported by the National Science Foundation, by the David & Lucile Packard Foundation, and by the Center for Subsurface Sensing and Imaging Systems (CenSSIS), an NSF Engineering Research Center.

# Figure Captions

Figure 1: Single-photon imaging. A single-photon source S emits a photon from position $\mathbf{x}$ and sends it through an optical system described by its impulse response function, $h(\mathbf{x}_1, \mathbf{x})$. A scanning single-photon detector D detects the photons and thus measures the probability density of photon arrivals, $p(\mathbf{x}_1)$. An unknown object (shaded region) is imbedded in the system.

Figure 2: Two-photon imaging. A two-photon source S emits one photon from point $\mathbf{x}$ and sends it through the system $h_1(\mathbf{x}_1, \mathbf{x})$ with an unknown object imbedded in the system, and emits the other from $\mathbf{x}'$ and sends it through $h_2(\mathbf{x}_2, \mathbf{x}')$. Two scanning detectors $D_1$ and $D_2$ record the singles, $p_1(\mathbf{x}_1)$ and $p_2(\mathbf{x}_2)$, and the coincidence rate $p(\mathbf{x}_1, \mathbf{x}_2)$. The shaded region is an imbedded object.

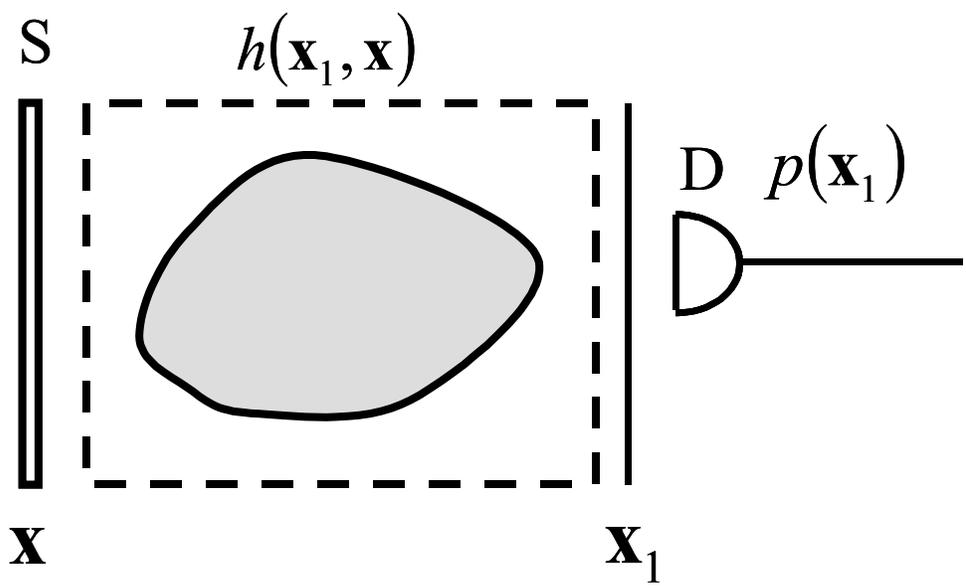

Fig. 1

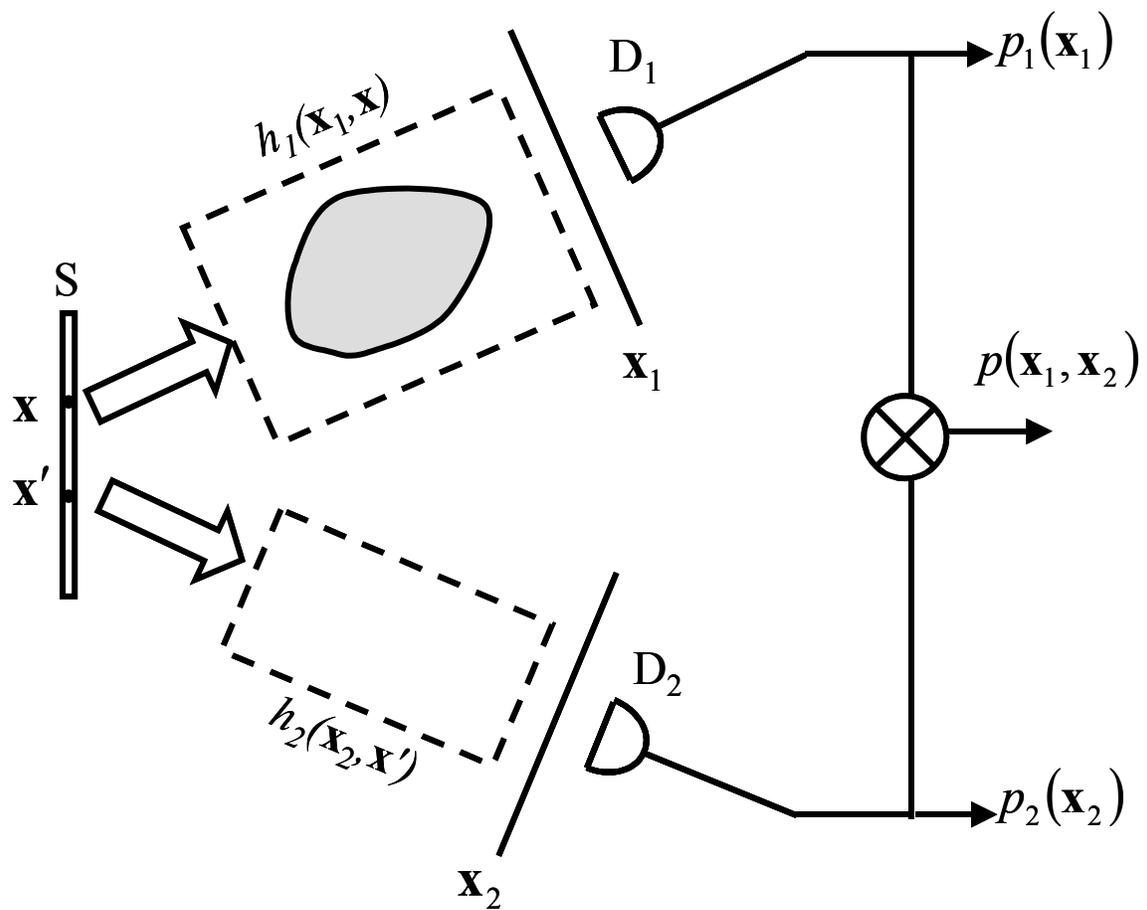

Fig. 2